\documentclass{article}
\begin{document}
\title{\bf Concept of unified local field theory\\ and nonlocality of matter}
\author{A.A. Chernitskii
\\ {\normalsize St.-Petersburg Electrotechnical University}
\\{\normalsize Prof. Popov str. 5, Russia, 197376;\quad aa@cher.etu.spb.ru}
}
\date{}
\maketitle

\begin{abstract}
The concept of unified local field theory is considered.
According to this concept the quantum description and the classical one must be
the levels for investigation of some world solution of the
unified field model. It is shown that
in the framework of the unified local field theory
there are nonlocal correlations between
space separate events. Thus the experiments of Aspect type for testing
of the Bell inequalities and for showing of the nonlocal correlations
do not reject a possibility for description of matter with
the unified local field theory. Advantages of
such theory for new technologies are considered.
\end{abstract}

\section{Historical introduction}

The whole history of pre-quantum physics naturally led to
the idea of unified field theory for description of matter.
All particles of matter and their
apparent mutual influence must be represented by some solution of
the appropriate field model which must be nonlinear.
Also this model must be local, i.e. it is represented by
some purely differential system of equations.
This is an essence of the ideas which was inspiring
for many scientists in their working.
Let us mention just a few: A.~Einstein, L.~de Broglie, H.~Weil, A.~Eddington,
G.~Mie, E.~Schr\"o\-din\-ger, M.~Born, L.~Infeld, J.~Plebansky, etc.

However, the impressive success of quantum mechanics
has eclipsed the idea of unified field theory which was in the air.
The quantum mechanics is essentially linear theory which is
easier for investigation than nonlinear one. But
the quantum mechanics gives the probabilistic predictions only.

Now opinions of physicists on the question about
fundamental character of the quantum mechanical description are divided.
One part believe that the indeterministic character of quantum mechanics
reflects some fundamental quality of nature. Other part believe
that this indeterminacy comes from an incompleteness
of the appropriate description for matter,
such as we have the indeterminacy in statistical mechanics.

\begin{figure}
\unitlength 1.5mm
\begin{picture}(69.67,33)
\put(5,0){
\begin{picture}(69.67,33)
\thinlines
\put(5,5){\vector(3,1){65}}
\put(35.00,15.00){\circle*{1.33}}
\put(10.00,6.67){\circle*{1.33}}
\thicklines
\put(35.00,15.00){\vector(3,1){7}}
\put(35.00,15.00){\vector(-3,-1){7}}
\put(60.00,23.33){\circle*{1.33}}
\put(60.00,23.33){\vector(0,-1){7}}
\put(60.00,23.33){\vector(-1,0){7}}
\put(10.00,6.67){\vector(0,1){7}}
\put(10.00,6.67){\vector(1,0){7}}
\put(19.67,12.00){\makebox(0,0)[cc]{$l$}}
\put(47,21){\makebox(0,0)[cc]{$l$}}
\put(9.00,3.33){\makebox(0,0)[cc]{$A$}}
\put(19.5,4.33){\makebox(0,0)[cc]{${\bf J}^{\prime}_A$}}
\put(7.0,15.33){\makebox(0,0)[cc]{${\bf J}^{\prime\prime}_A$}}
\put(60.67,26.5){\makebox(0,0)[cc]{$B$}}
\put(63.50,16.67){\makebox(0,0)[cc]{${\bf J}^{\prime\prime}_B$}}
\put(51.67,26.33){\makebox(0,0)[cc]{${\bf J}^{\prime}_B$}}
\put(37,12.50){\makebox(0,0)[cc]{$O$}}
\thinlines
\put(35.00,5.00){\vector(0,1){20}}
\put(25.00,15.00){\vector(1,0){20}}
\put(45.00,13){\makebox(0,0)[cc]{$x^1$}}
\put(33.3,25.50){\makebox(0,0)[cc]{$x^2$}}
\put(69.67,24){\makebox(0,0)[cc]{$x^3$}}
\end{picture}
}
\end{picture}
\caption{Towards an experiment for demonstration of nonlocal correlations.}
\label{Fig:nonlocexp}
\end{figure}
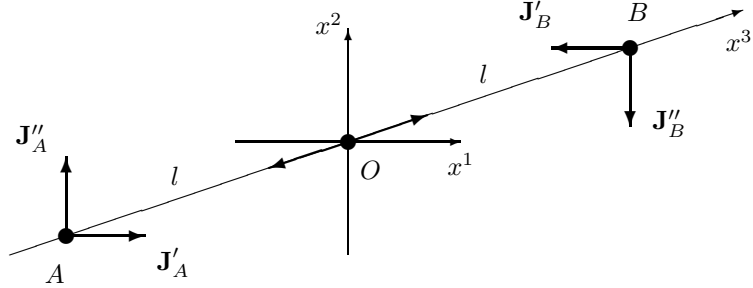

Einstein, Podolsky, and Rosen in their famous article \cite{EPR}
advanced the arguments for the standpoint that quantum mechanical
description of reality is incomplete. Here we repeat the essence of
the EPR paradox for the example which was given by Bohm and Aharonov
\cite{Bohm&Aharonov1957}.

Let us consider the experimental scheme (see figure \ref{Fig:nonlocexp}),
in which a source (at the point $O$) gives birth to the pair one-half spin
particles from a zero spin state.
Then these particles move in opposite directions (along the axis $x^3$).
Travelling considerable distance ($l$), they come to
detectors (at the points $A$ and $B$), which determine
the spin states of the particles.
Stern-Gerlach magnets can be used as these detectors which finally
determine the vector of angular momentum for the particles
(${\bf J}_A$ and ${\bf J}_B$).
Suppose we can arbitrarily take the axis $x^1$ or $x^2$
for orientation of the magnets to measure the appropriate
projection of spin.

According to the quantum mechanical description for this experiment,
the spin states of the individual particles
are indeterminate until the measurement event.
But as soon as we have measured the spin state for one particle
then the spin state for another particle becomes determinate
immediately. This resulting situation is connected with
the conservation law
of full angular momentum for the system of two particles.

Let an orientation of the magnet (along the axis $x^1$ or $x^2$) at the point $A$
be chosen by a chance switch. Now if
the detector at the point $A$ gives the value of angular momentum
${\bf J}^\prime_A$ or ${\bf J}^{\prime\prime}_A$
(at figure \ref{Fig:nonlocexp}) then
another detector immediately gives
${\bf J}^\prime_B$ or ${\bf J}^{\prime\prime}_B$
respectively. Because we can take the distance $l$ as arbitrary long,
this situation looks as contradiction with the
thesis for locality of interactions.
Thus quantum mechanics predicts nonlocal correlations between the events.
Well known Aspect  experiment \cite{Aspect}  for testing of
also well known Bell inequalities \cite{Bell}
determines that there are the nonlocal correlations.

At first glance the existence of this nonlocal correlations
rejects a possibility for description of matter by
the unified local field theory. However, actually, this is not the fact.
In the following section we show that the nonlocal correlations
between the events must exist in the unified local field theory of matter.

\section{Fundamental concept of matter}

Actually the concept of unified local field theory for the material world
is similar to the concept of ether, if we understand it
in the broad sense but not narrow mechanical one.
This concept supposes only two basic properties: continuity and
locality. Mathematically these properties are expressed in the fact that
we consider some purely differential field model or
some system of equations with partial derivatives.
To describe naturally the interactions between material objects,
this system of equations must be nonlinear.
Also we believe that there is a model solution which is determinate
in the whole three-dimensional space at each point of time.
Thus, according to this concept, we can
consider some Cauchy problem or the problem with initial and boundary
condition for obtaining the world evolution.

Within the framework of such theory a single elementary particle
is represented by some space-localized solution.
Moreover, because
elementary particles have wave
properties, this solution must have the appropriate wave part.
The wave part is considered here in the sense of time Fourier expansion
for the solution in own coordinate system of the particle,
where this part has the form of standing wave.

There is the simplest example for such standing wave
even for customary linear wave equation. These well known solutions
of the wave equation in spherical coordinate system include
spherical harmonics. For the spherically symmetric case
we have the standing wave
\begin{eqnarray}
\nonumber
\frac{\sin (\underline{\omega}\, r)}{\underline{\omega}\, r}\,
\sin (\underline{\omega}\, x^0 )
\end{eqnarray}
which is formed by the sum of divergent and convergent
spherical waves. With the help of Lorentz transformation we can obtain
the appropriate solution in the form of moving
nondeliquescent wave packet.
Then the own frequency
$\underline{\omega}$ transforms to wave vector $k_\mu$ such that
\mbox{$|k_\mu\,k^\mu| {}={} \underline{\omega}^2$}.

A single elementary particle solution of a nonlinear field
model may be called also as {\em solitron}. This term has a similar sense that
``solitary wave'' or ``soliton''. But usually the term ``soliton''
is used in mathematical context for the case of special solutions.

It is significant,
the concept of unified field theory supposes that
all variety and evolution of the material world
are represented by some space-time field
configuration which is an exact solution of the nonlinear field model.
It is evident that this solution is very very complicated
but it is determinate in space-time by the field model
with initial and boundary conditions.
In the vicinity of a separate elementary particle this world solution
is close to the appropriate single elementary particle solution,
but each elementary particle behaves as the part of the world solution.
Thus the behavior of each elementary particle is connected
with the whole space-time field configuration for the world solution.

For certain conditions it is possible to consider
the world solution part connecting with the separate elementary particle
as the appropriate solitron solution with slowly variable velocity.
(For the case of nonlinear electrodynamics see,
for example, the article \cite{Chernitskii1999}.)
This level for investigation of the world solution relates to the
classical (not quantum) physics.

It is evident that although the model is local, the world solution
is nonlocal in character because it is determined on the whole space-time
applicable domain.
This means, in particular, that there are undoubtedly
nonlocal correlations between space separate parts
of the common world solution.
This sentence may be explained with the help of the following simplest
example.

Let us consider the customary plane wave on axis $x^3$
with fixed wave-length $\lambda$.
Let this wave be the solution of the customary
linear wave equation and at the point $O$ with coordinates
\mbox{$(x^1,x^2,x^3) {}={} (0,0,0)$}
the field evolution has the form \mbox{$a\sin [(2\pi/\lambda)\,x^0]$}.
Then at the point $Q$ \mbox{$(0,0,q)$}
the field evolution has the form
\mbox{$a\sin [(2\pi/\lambda)\,(x^0 {}-{} q)]$}
and at the point $P$ \mbox{$(p,0,0)$} the field evolution
is the same that at the point $O$.
Thus here there is the nonlocal correlation between the field
evolution at the points $O$, $Q$, and $P$.
Totality of such nonlocal correlations is, in fact, the solution
in space-time for local field model. The possible world solution
(which is extremely more complicated than the plane wave)
is also the continuous set of nonlocal correlations for the field
evolution at the points of three-dimensional space.

Of course, if we make some excitation for field at the point
$O$ then a propagation of this excitation from this point
will have a finite speed. But in the scope of the world
solution we
are not able
to make this excitation or
to modify arbitrarily this world solution.
Any excitations of the field at the point $O$
belong to the world solution which is a single whole.
That is, in this case we must consider also all excitations
coming to this point and we will have some standing wave near it.
Thus the world solution is rather a very complicated system of
standing waves than progressing ones.
The initial condition is a common cause of all field excitations
and after a long evolution the different correlations may exist,
even the strange ones.
It can only be said quite positively that the world solution
can be represented by Fourier integral (or series) on orthogonal
space-time harmonics which are essentially nonlocal.
(Here we must remember how
a dominant role is played by orthogonal functions in quantum approach.)

Let us consider once more the example
with two particles scattering in the opposite directions,
shown on figure \ref{Fig:nonlocexp}.
According to the concept under consideration there is the
appropriate two-particle or two-solitron solution of the field model.
Of course, according to this approach,
in reality
there is only the world solution but in our case
we have the experimental scheme which is prepared specially for
investigation of some aspects of the world solution part approximated by
the two-particle solution.
In particular, this solution must satisfy the conservation law
of full angular momentum. Just this is confirmed by the result of
the experiment.
It is obvious that the magnitude of distance between the detectors
 is not essential here.

The key to understanding the appearance of momentary distant interaction
in this experiment is contained in the concept of chance choice.
Within the framework of the world solution a chance choice is absent,
but both experimenter and experimental apparatus are a part of this
world solution.
That is the orientation of particle spin detectors in the experiment
under consideration is predetermined by the world solution.
We speak about
the chance choice because we do not know the world solution.

As experimentalists, we think that we establish the initial conditions
for the process under investigation but may be this is
too conceitedly and the veritable initial condition is established
earlier. But as theorists, we can already calculate many
correlations between the space-time events.

Thus we can suppose that the quantum mechanical description is the level
for investigation of the world solution.
This level takes into consideration, in particular,
the global or nonlocal aspects of this solution.

Nonlocality was founded in quantum mechanics from the outset.
In Schr\"o\-din\-ger's picture a free elementary particle
(which has a determinate momentum) is related with
 a plane wave having a constant amplitude on the whole space.
In this case the
quantum mechanical description does not determine a position of
the particle. That is we have the representation of the free elementary
particle by non space-localized wave that accentuates just nonlocal
aspect of matter.

As we see, there is nonlocality also in the framework of unified
local field theory. But such theory supposes a solitron model for a free
elementary particle which is intuitively more preferable. Furthermore,
according to this concept  there is the deterministic description of matter.

The separate question is that the world solution concept excludes a
free will for a person. But we can suppose that
a possible will agent is outside of space-time world solution.
This will agent can be called an individual spirit which is a part of some
spiritual world.
We can also suppose that the individual spirit can partially modify
the world solution using some dynamical boundary conditions. But these
modifications must be nonlocal in general.

 Nowadays an educated individual knows that there are the laws of nature
 but he believes that the initial conditions may be established by
an independent deliberate action. (A man sets free a massive body and it
rushes to the earth with a constant acceleration but
the man chooses the space-time point for beginning of the movement.)
If we accept this point of view then we must assume that
there is something outside the material world and it realizes the
choice.
This choice might be realized by the individual spirit.
But a possible border between
the material world and the spiritual one is fuzzy. Some philosophical
systems suppose even that a powerful spirit can modify the laws
of the material world because it is constructed by consciousness.
In connection with this topic we should remember also the discussion
between A.~Einstein and R.~Tagore \cite{Einstein&Tagor}.

In any case we have the repeatable
experimental confirmations for existence
of the material world laws and these experiments do not exclude
the description based on the concept of unified local field theory.
In particular, nonlinear electrodynamics of Born-Infeld type
with dyon singularities \cite{Chernitskii1999}
(see also \cite{Chernitskii1998HPA,Chernitskii1998JHEP,%
Chernitskii2000Bidyon,Chernitskii2000BICliff}) may be considered
in this connection. In this theory there is a field configuration,
named bidyon, which can be a model for a particle with spin.

\section{Prospects for applications}

At present we consider a single atom and even a single electron
as objects of technology. There is a concept of a single electron
transistor \cite{SET} and we can seriously consider prospects
for building an Avogadro-scale computer
acting on \mbox{$\sim 10^{23}$} bits \cite{LloydNature2000}.
In such computer using
the nuclear magnetic resonance one nuclear spin must store one bit
of information (see article \cite{LloydNature2000} and
the references contained therein).

Traditional computation can do many useful things and this ability
can become very much stronger with the possible
Avogadro-scale technology. But the traditional computation
needs a determinate controlling. Such controlling is possible if
we have the unified field theory of matter in the sense that was stated
above.

This is one of the possible applications of the approach under review.
But, of course, a realization for the paradigm of
unified field theory will discover
abilities which we do not know at the present time.

\end{document}